\documentclass{Interspeech-Paper-Kit/Interspeech}

\interspeechcameraready

\usepackage[hang,flushmargin]{footmisc} 
\usepackage{graphicx}
\usepackage{subcaption}
\usepackage{verbatim}
\usepackage{booktabs}
\usepackage{siunitx}
\usepackage{makecell}
\usepackage{microtype}

\title{Deep learning based spatial aliasing reduction in beamforming for audio capture}

\author[affiliation={1}]{Mateusz}{Guzik}
\author[affiliation={2}]{Giulio}{Cengarle}
\author[affiliation={2}]{Daniel}{Arteaga}

\affiliation{Institute of Electronics}{AGH University of Krakow}{Poland}
\affiliation{Dolby Laboratories}{Barcelona}{Spain}

\email{mateusz.guzik@agh.edu.pl, giulio.cengarle@dolby.com, daniel.arteaga@dolby.com}


\begin{document}

\maketitle

\newcommand{\brc}[1]{\ensuremath{\left(#1\right)}} 
\newcommand{\crBrc}[1]{\ensuremath{\left\{#1\right\}}} 
\newcommand{\sqBrc}[1]{\ensuremath{\left[#1\right]}} 
\newcommand{\vtBrc}[1]{\ensuremath{\left|#1\right|}} 
\newcommand{\tps}[1]{\ensuremath{#1^\mathrm{T}}} 
\newcommand{\nrm}[1]{\ensuremath{\lVert#1\rVert}} 
\newcommand{\scl}[1]{\ensuremath{#1}}
\newcommand{\vct}[1]{\ensuremath{\mathbf{#1}}}
\newcommand{\mtr}[1]{\ensuremath{\mathbf{#1}}}
\newcommand{\ten}[1]{\ensuremath{\mathbf{#1}}}
\newcommand{\freqIdx}{\ensuremath{f}} 
\newcommand{\freqNum}{\ensuremath{F}} 
\newcommand{\enFreqInd}{\ensuremath{\freqIdx = 1, \ldots, \freqNum}} 

\newcommand{\frameIdx}{\ensuremath{t}} 
\newcommand{\frameNum}{\ensuremath{T}} 
\newcommand{\enFrameInd}{\ensuremath{\frameIdx = 1, \ldots, \frameNum}} 

\newcommand{\micIdx}{\ensuremath{i}} 
\newcommand{\micNum}{\ensuremath{I}} 
\newcommand{\enMicInd}{\ensuremath{\micIdx = 1, \ldots, \micNum}} 

\newcommand{\srcIdx}{\ensuremath{s}} 
\newcommand{\srcNum}{\ensuremath{S}} 
\newcommand{\enSrcInd}{\ensuremath{\srcIdx = 1, \ldots, \srcNum}} 

\newcommand{\shdIdx}{\ensuremath{l}} 
\newcommand{\shdNum}{\ensuremath{L}} 
\newcommand{\enShdInd}{\ensuremath{\shdIdx = 1, \ldots, \shdNum}} 

\newcommand{\dirIdx}{\ensuremath{d}} 
\newcommand{\dirNum}{\ensuremath{D}} 
\newcommand{\enDirInd}{\ensuremath{\dirIdx = 1, \ldots, \dirNum}} 

\newcommand{\chIdx}{\ensuremath{c}} 
\newcommand{\chNum}{\ensuremath{C}} 
\newcommand{\enChInd}{\ensuremath{\chIdx = 1, \ldots, \dirNum}} 
\newcommand{\vrtMicVec}{\ensuremath{\vct{v}_{\freqIdx \frameIdx}}} 
\newcommand{\numVrtMic}{\ensuremath{\scl{V}}}
\newcommand{\fltMat}{\ensuremath{\mtr{F}_{\freqIdx \frameIdx}}} 
\newcommand{\decMat}{\ensuremath{\mtr{D}}} 
\newcommand{\decVec}{\ensuremath{\vct{d}_{\freqIdx \frameIdx}}} 
\newcommand{\trgVec}{\ensuremath{\vct{t}_{\freqIdx \frameIdx}}} 
\newcommand{\trgEncMat}{\ensuremath{\mtr{E}}} 
\newcommand{\srcVec}{\ensuremath{\vct{s}_{\freqIdx \frameIdx}}} 
\newcommand{\decSrcAng}{\ensuremath{\theta_{\dirIdx \srcIdx}}} 
\newcommand{\shpCoe}{\ensuremath{\scl{\alpha}}} 
\newcommand{\trEncScl}{\ensuremath{\scl{e}}} 
\newcommand{\decDir}{\ensuremath{\vct{i}_\dirIdx}} 
\newcommand{\srcDir}{\ensuremath{\vct{j}_\srcIdx}} 
\newcommand{\alsFreq}{\ensuremath{\scl{\bar{f}}}} 
\newcommand{\micSpac}{\ensuremath{\scl{d}}} 
\newcommand{\sndSpd}{\ensuremath{\scl{c}}} 
\newcommand{\cmpSet}{\ensuremath{\mathbb{C}}}
\newcommand{\realSet}{\ensuremath{\mathbb{R}}}
\newcommand{\idtMat}{\ensuremath{\mtr{I}}}
\newcommand{\mskMat}{\ensuremath{\scl{M}_{\chIdx \freqIdx \frameIdx}}} 
\newcommand{\dani}[1]{{\color{red}DANI: #1}}

\begin{abstract} 
Spatial aliasing affects spaced microphone arrays, causing directional ambiguity above certain frequencies, degrading spatial and spectral accuracy of beamformers. Given the limitations of conventional signal processing and the scarcity of deep learning approaches to spatial aliasing mitigation, we propose a novel approach using a U-Net architecture to predict a signal-dependent de-aliasing filter, which reduces aliasing in conventional beamforming for spatial capture. Two types of multichannel filters are considered, one which treats the channels independently and a second one that models cross-channel dependencies. The proposed approach is evaluated in two common spatial capture scenarios: stereo and first-order Ambisonics. The results indicate a very significant improvement, both objective and perceptual, with respect to conventional beamforming. This work shows the potential of deep learning to reduce aliasing in beamforming, leading to improvements in multi-microphone setups.

\end{abstract}

\newcommand\blfootnote[1]{%
  \begingroup
  \renewcommand\thefootnote{}\footnote{#1}%
  \addtocounter{footnote}{-1}%
  \endgroup
}
\blfootnote{
This work was done during Mateusz Guzik's internship at Dolby Laboratories.
}


\section{Introduction}

Beamforming enables the extraction of signals from desired directions using an array of microphones \cite{benesty2008microphone}.
While often used as a preprocessing step, such as denoising and separation for automatic speech recognition, beamforming plays a central role in spatial audio, where it enables audio encoding, e.g.\ obtaining directional stereo signals from a spaced pair of omnidirectional microphones \cite{faller2010conversion} or converting a multi-microphone recording into Ambisonics \cite{zotter2019ambisonics}.

Similarly to how the discrete sampling interval in time domain determines the frequency limit above which time domain aliasing occurs, spatial sampling is also subject to an analogous limitation, known as spatial aliasing \cite{benesty2008microphone, dmochowski2008spatial, rafaely2007spatial}. 
Spatial aliasing affects spaced microphone arrays, where processing based on phase differences suffers from directional ambiguity above the spatial aliasing frequency, drastically degrading spatial and spectral accuracy of beamformers.

Considering conventional signal processing, efforts to mitigate spatial aliasing in beamforming have focused on wide-band and multi-stage solutions \cite{zhang2024anti, tang2011aliasing, 6868957}, rotating arrays \cite{cigada2008moving} and modeling of spatial aliasing for spherical arrays \cite{4538672, brown2019spatial, 7331617}.
Despite significant interest in deep learning approaches to signal processing, the topic of spatial aliasing mitigation has so far attracted little attention in the deep learning audio community, often being approached indirectly.
To the best knowledge of the authors, currently there is no literature that explicitly targets the reduction of spatial aliasing with deep learning, specifically in the context of spatial audio.
Examples of indirect approaches to the reduction of spatial aliasing include neural beamforming  \cite{gu2022towards}, multi-channel source separation \cite{10321676, herzog2023ambisep,  wang2018combining}, neural Ambisonic encoding  \cite{heikkinen2024neural}, and Ambisonic upmixing \cite{xia2023upmix}; in most of these cases the training objectives indirectly aim at the restoration of the target signal above the spatial aliasing frequency.
In such systems, the de-aliasing capabilities are not directly evaluated, as their responses are a composite of multiple simultaneous enhancement objectives.
In addition, neural beamforming and source separation systems are in general not directly applicable to sound field decomposition.

In this work, we consider the reduction of spatial aliasing in beamforming, using a deep learning approach.
In particular, we propose a U-Net based filter, that reduces the effects of spatial aliasing in the signals obtained as the result of sound field encoding and decoding.
Our approach allows to directly evaluate the de-aliasing capabilities of the system and has the potential to be easily integrated into existing spatial audio processing applications, e.g.\ obtaining first-order Ambisonics (FOA) signals from a tetrahedral microphone array, or obtaining directional stereo signals from compact omnidirectional arrays.

This paper is structured as follows: in Section 2, the effects of spatial aliasing on audio capture are discussed; Section 3 introduces the proposed filtering model; in Section 4, the filter estimation procedure is described; Section 5 contains the details of the experimental evaluation, results and discussion; Section 6 presents final remarks and conclusions.
\section{Aliasing in spatial audio capture}\label{sec:spatial_aliasing}
Spatial processing of audio signals captured with spaced microphones is in general based on phase differences between the sensors, leveraging the property that in a certain frequency range they map univocally to the Direction of Arrival (DoA) of an acoustic wavefront \cite{benesty2008microphone}. 
The upper frequency limit of the aforementioned range depends on the spacing between the microphones, such that
\(
    \alsFreq
    =
    \sndSpd
    \ 
    /
    \brc{2 \micSpac}
\)
, where \alsFreq{} is the spatial aliasing frequency, \sndSpd{} is the speed of sound and \micSpac{} is the spacing between a pair of microphones.
Above \alsFreq{}, the phase differences do not unambiguously map to the DoA, because sound originating from different directions can cause identical shift through phase wrapping.
For a comprehensive discussion on spatial aliasing, the reader is kindly referred to \cite{benesty2008microphone, dmochowski2008spatial, rafaely2007spatial}. 

\begin{figure}[!h]
    \centering
    \includegraphics[width=.6\linewidth]{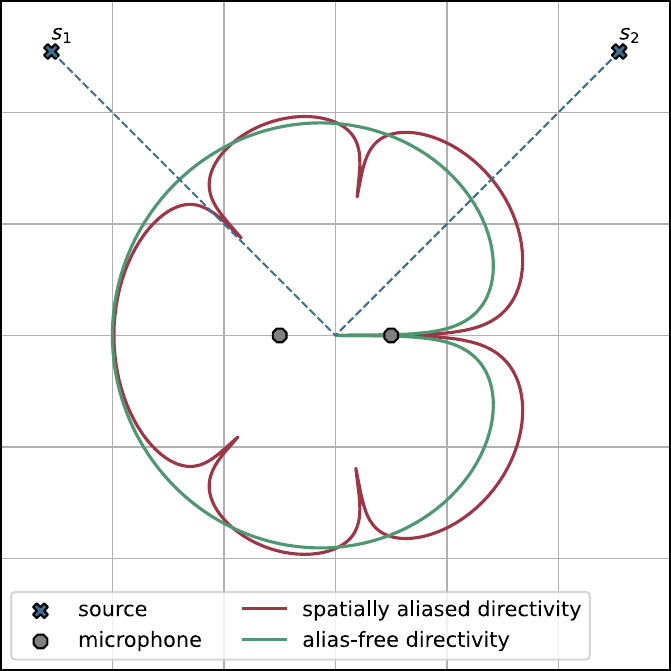}
    \caption{Polar representation of spatially aliased conventional beamforming. The curves are in arbitrary logarithmic scale.}
    \label{fig:spatial_aliasing}
\end{figure}

Figure \ref{fig:spatial_aliasing} presents the effects of aliasing on spatial capture using conventional beamforming, where two omnidirectional microphones are used to form a cardioidal response at a frequency above \alsFreq{}. For each curve, the distance from the center of the microphone array represents the magnitude response for the corresponding direction.
In this example, the spatial response is deformed due to spatial aliasing.
As a result, the signal of source
\(
    s_1
\)
is improperly attenuated, while the signal of source
\(
    s_2
\)
is improperly amplified, with respect to the alias-free cardioid directivity.
Since these alterations caused by spatial aliasing are frequency-dependent, source signals will be subject to direction-dependent filtering, while in the alias-free case they would be simply scaled by frequency-independent directional factors.
This implies that in case of spatial capture, where the goal is to obtain a specified frequency-independent spatial response, a frequency- and direction-dependent combination of both attenuation and amplification is needed to reduce spatial aliasing with a filtering approach.
On the other hand, for source separation, the beamformer is steered towards the direction of the target source and the goal is to obtain the greatest possible spatial selectivity, rather than a smooth amplitude response across all directions.
\section{Proposed de-aliasing filtering}\label{sec:section_3}

In this work, we address the de-aliasing of beamformed signals, which we will refer to as \emph{virtual microphones}, in the Short-Time Fourier Transform (STFT) complex domain.  

Specifically, we formulate the de-aliasing process for a given time-frequency tile of  \numVrtMic{} virtual microphone signals arranged as a column vector,
\(
    \vrtMicVec{} \in \cmpSet^{\numVrtMic \times 1},
\)
with \frameIdx{} being a frame index and \freqIdx{} being a time index.
The de-aliased signals
\(
    \decVec \in \cmpSet^{\dirNum \times 1}
\)
can be obtained from the aliased signals though the following equation:
\begin{equation}\label{eq:filtering_model}
    \decVec
    =
    \decMat
    \ 
    \fltMat
    \ 
    \vrtMicVec
    ,
\end{equation}
where
\(
    \fltMat \in \cmpSet^{\numVrtMic \times \numVrtMic}
\)
is the predicted de-aliasing filter matrix, which is the main concern of our approach, and
\(
    \decMat \in \realSet^{\dirNum \times \numVrtMic}
\)
is a pre-defined decoder matrix, discussed below.
In this paper, we consider only frequency-independent decoding, although equation \eqref{eq:filtering_model} could be easily generalized by making the decoding matrix frequency-dependent.

The pre-defined matrix \decMat{} in equation \eqref{eq:filtering_model} decodes the virtual microphone signals into a specific reproduction setup.
Examples include the mid-side decoding \cite{johnston1992sum} and the Ambisonic decoding \cite{zotter2019ambisonics}. 
When the virtual microphone signals correspond directly to the desired decoded signals,  e.g.\ for two left- and right-facing cardioid beamformers to be rendered in stereo, no decoding is necessary. In this case, \decMat{} reduces to an identity matrix 
\(
    \idtMat^{\numVrtMic \times \numVrtMic}
\)
and \decVec{} corresponds directly to the de-aliased virtual microphone signals. Otherwise, \decVec{} corresponds to the de-aliased and decoded signals.
This approach is inspired by \cite{usat_2024} and addresses the fact that certain decoding strategies may result in a loss of spatial information compared to the virtual microphone signals. For instance, decoding Ambisonics to stereo typically discards the vertical component.
In such cases, it proves advantageous to emphasize de-aliasing of the sound field components that dominate the decoded signals, while reducing emphasis on less significant components.
However, our approach is generic enough so that it can achieve both alias-free decoding and alias-free virtual microphones, depending on the adopted matrix \decMat{}.


Note that the de-aliasing filter acts on the virtual microphones, rather than directly on the decoded signals. By using a multichannel matrix formulation where the virtual microphones can interact, the system is given the added capacity of leveraging a plurality of virtual microphones to de-alias each individual virtual microphone, effectively performing an implicit intermediate beamforming step.
However, if intermediate beamforming is to be prevented, the filter can be reduced to a diagonal matrix.
In the remaining parts, we refer to the former filter setting as \textit{\textbf{full}} and to the latter as \textit{\textbf{diag}}.
The general formulation of equation \eqref{eq:filtering_model} provides a flexible way to encapsulate a variety of possible scenarios in one model; as an example, if the goal is to produce alias-free non-decoded Ambisonics, the training can be augmented with different decoders and decoding positions, potentially increasing the system robustness and indirectly enabling de-aliasing on the full sphere.
\section{U-Net-based filter prediction}\label{sec:section_4}
We propose to predict the de-aliasing filter matrix \fltMat{} based on the virtual microphone signals \vrtMicVec{}, leveraging supervised training of a deep neural network, similarly to \cite{heikkinen2024neural}.
Since the virtual microphone signals are characterized by different spatial responses or looking directions, as in the case of different Ambisonic components, spatial aliasing will affect each virtual microphone in a different way. This useful diversity and redundancy makes virtual microphone signals a suitable regressor for the proposed filter-predicting model.


\subsection{Alias-free supervisory signals}

Assuming \srcNum{} anechoic source signals \srcVec{} and an alias-free encoder matrix $\trgEncMat{} \in
    \realSet^{\dirNum \times \srcNum}$, we construct the target alias-free decoded beamformer signals 
\(
    \trgVec
    \in
    \cmpSet^{\dirNum \times 1}
\)
using the equation:
\begin{equation}
    \trgVec
    =
    \trgEncMat
    \ 
    \srcVec
    .
\end{equation}
For supervised learning,
 \trgVec{} is the target for \decVec{} in the sound field generated by the sources \srcVec{}.

In this work, we focus on a frequency-independent alias-free encoder matrix \trgEncMat{} corresponding to a first-order microphone:
\begin{equation}\label{eq:polar_patterns}
        \sqBrc{
        \trgEncMat}_{
        \dirIdx
        \srcIdx}
    =
        \shpCoe
    +
    \brc{
        1
        -
        \shpCoe
        }
        \tps{
            \decDir
        }
        \ 
        \srcDir,
\end{equation}
where \shpCoe{} is a shape coefficient and \decDir{}, \srcDir{} are directional unit vectors in Cartesian coordinates.
The directional vector \decDir{} points towards the \dirIdx{}-th decoding direction, realized by the \dirIdx{}-th row of the decoder matrix \decMat{}, while \srcDir{} points towards the \srcIdx{}-th sound source. \shpCoe{} is a coefficient between 0 and 1 that defines the shape of the spatial response, such that 0 results in a figure-of-eight pattern, 0.5 produces a cardioid and 1 corresponds to an omnidirectional characteristic.
The application of the alias-free encoder matrix \trgEncMat{} to the source signal vector \srcVec{} can be illustrated using Figure \ref{fig:spatial_aliasing}, where the intersection of the dashed source DoA lines with the target spatial directivity determines the values for the weighted summation.
Note that \trgEncMat{} does not model the direction-dependent phase shifts potentially introduced by a beamformer.
Therefore, the complex-valued filter \fltMat{} will also be tasked with relative phase restoration, approximating a spatially uniform phase response of the de-aliased signals.

\subsection{Model, data and training procedure}\label{sec:training}
Following state-of-the-art neural processing of audio \cite{gul2023survey},  the U-Net \cite{ronneberger2015u}, an encoder-decoder architecture based on 2D convolutional filters, is employed as the filter-predicting model.
The only difference between the adopted model and the original U-Net is the final activation function, which in our case is an identity.
This choice was dictated by the required complex-valued filter properties, as described in Section \ref{sec:section_3}.

Similarly to \cite{heikkinen2024neural}, real and imaginary parts of the virtual microphone signals are used as an input to the model, represented by real-valued features unfolded across the channel dimension.
Based on this representation, the model predicts a set of time-frequency masks
\(
    \mskMat \in \realSet
    ,
\)
where \chIdx{} is the channel index,  which are then reshaped to create the filter matrix \fltMat{}.
As discussed in Section \ref{sec:section_3}, the filter \fltMat{} can be either diagonal, with off-diagonal terms zeroed out, or it can be a full matrix filter that includes the cross-terms.
Depending on the filter configuration, the number of output channels \chNum{} is either equal to \(2\numVrtMic{}\) or
\(
    2\numVrtMic^2
    ,
\)
respectively, where the factor 2 accounts for the aforementioned real-valued representation of the complex-valued masks.

The PHASEN \cite{yin2020phasen} loss between the de-aliased \decVec{} and the alias-free \trgVec{} decoded signals is optimized by minimizing its mean value across the decoding directions.
The model is trained for 100 epochs using the Adam optimizer with 
\(
    \beta_1 = 0.9
    ,
    \beta_2 = 0.999
    ,
\)
a batch size of 24 and a learning rate of 0.001, which is halved if the batch loss does not decrease for two epochs.

The training is based on simulated microphone array signals obtained by applying delays to source signals, based on anechoic far field propagation with respect to the center of the array.
We simulate a 4-channel omnidirectional microphone array, with the sensors arranged on a cross-like plan, where one pair is aligned with the x-axis and the second pair is aligned with the y-axis.
In the experimental setups with a fixed aliasing frequency, the microphone spacing \micSpac{} for each pair is equal to \SI{3}{\cm}, while in the case of varying aliasing \micSpac{} is chosen randomly between 1 and \SI{10}{\cm}, independently for each pair.
Microphone mixture signals are obtained by choosing between one and four sources, simulating the capsule signals with the sources randomly placed around the array and summing their contributions for each sensor.
The signals are then transformed by an STFT with 2048-point Hann window and 1024 overlapping samples.

As source signals we use \SI{5}{\s} excerpts of speech, music, noise and environmental signals, sampled at \SI{44.1}{\kHz}.
The speech samples are randomly drawn from \cite{openslr} and additional 300 hours of speech recordings.
The music signals are randomly drawn from a set of 4000 songs, while the noise and environmental recordings are randomly drawn from \cite{chen2020vggsoundlargescaleaudiovisualdataset, fonseca2022fsd50kopendatasethumanlabeled}.
\section{Experimental evaluation}

We investigate two experimental setups: \textit{\textbf{i)}} two opposite-facing cardioids and \textit{\textbf{ii)}} 2D First-Order Ambisonics, aka.\ planar, horizontal or circular FOA \cite{zotter2019ambisonics}.
Additionally, both experiments are also subject to two variants, one with fixed and one with varying microphone spacing, which we refer to as \textit{\textbf{fix}} and \textit{\textbf{var}}.

Experiment \textit{\textbf{i)}} considers the case of two opposite-facing cardioid beams, a common setup for obtaining stereo signals from a pair of closely spaced omnidirectional microphones \cite{faller2010conversion}.
Given raw mixture recordings, the virtual microphone signals are obtained by calculating pressure gradient for the x-axis pair of the 4-channel microphone array described in Section \ref{sec:training}, which results in quasi-frequency-independent spatial characteristics.
In this experimental setup, we use an identity decoder, thus the spatial responses of the virtual microphone signals are explicitly optimized, as discussed in Section \ref{sec:section_3}.
To obtain suitable alias-free target signals, we follow the procedure described in Section \ref{sec:section_4}, with \shpCoe{} = 0.5 in equation \eqref{eq:polar_patterns}, while \numVrtMic{} = \dirNum{} = 2. 

Experiment \textit{\textbf{ii)}} addresses the de-aliasing of horizontal FOA, obtained from the aforementioned array of four microphones.
Similarly to experiment \textit{\textbf{i)}}, pressure gradient is calculated along the x- and y-axis pairs to obtain X and Y figure-of-eight signals respectively, while the omnidirectional component W is taken as an average of the four microphone signals.
In this experiment, \decMat{} is an in-phase decoder \cite{zotter2019ambisonics}, which produces four cardioids along the front, back, left, right directions, such that \numVrtMic{} = 3 and \dirNum{} = 4.

In both experiments, the raw mixtures are encoded to virtual microphone signals using conventional beamforming, based on calculation of the pressure gradient \cite{faller2010conversion}.
We have decided to use this type of beamformer to ensure a quasi-frequency-independent spatial response, while generalization is left as a future research possibility.
We report the results of both objective and subjective evaluation, described in the following sections. Additionally, we provide audio samples\footnote{\url{metlosz.github.io/dealiasing_audio_samples}}.

\subsection{Objective evaluation}
The objective evaluation is based on the Complex Scale-Invariant Signal-to-Noise-Ratio (C-Si-SNR) \cite{ni2021wpd++} between the de-aliased \decVec{} and the target decoded alias-free \trgVec{} signals, averaged across the \dirNum{} decoding directions.
Additionally, we heuristically calculate spatial responses for the trained model by simulating one source at a time for a grid of directions, predicting and applying the de-aliasing filter, and calculating the magnitude of the de-aliased signal.
The procedure is repeated with 100 different source signals for each grid direction, the magnitudes are aggregated and the results are normalized.

Table \ref{tab:SNR_results} reports the C-Si-SNR values in dB for experiments \textit{\textbf{i)}} and \textit{\textbf{ii)}} with fixed and varying microphone spacing.
Both experiments show an average C-Si-SNR close to \SI{30}{dB} for \textit{\textbf{fix}} and between \SI{15}{dB} and \SI{20}{dB} for \textit{\textbf{var}}, which is consistent with the expected increasing difficulty of restoring spatial aliasing with varying aliasing frequency.
In case of the 2D FOA experiment \textit{\textbf{ii)}}, where cross-microphone information is already utilized in the decoding step, the differences between \textit{\textbf{diag}} and \textit{\textbf{full}} seem negligible.
Regarding the cardioid pair experiment \textit{\textbf{i)}}, the \textit{\textbf{full}} filter shows an advantage over \textit{\textbf{diag}} for \textit{\textbf{var}} conditions, while for \textit{\textbf{fix}} the difference is less pronounced.
We hypothesize that this improvement could be explained by the re-introduction of the cross-channel interaction, which is not present in case of \textit{\textbf{i) var diag}}, neither in the filter nor in the decoder.


\begin{table}[h!]
\centering
\caption{C-Si-SNR values in dB for the experiments with \textit{\textbf{i)}} cardioid pair and \textit{\textbf{ii)}} 2D Ambisonics. The numbers in parentheses denote improvement over the aliased conditions.}
\label{tab:SNR_results}
\begin{tabular}{lcccc}
\toprule
 & \textit{\textbf{i) fix}} & \textit{\textbf{i) var}} & \textit{\textbf{ii) fix}} & \textit{\textbf{ii) var}} \\ 
\midrule
\textit{\textbf{diag}} & 
  \makecell{30.3$\pm$2.5 \\ (19.5)} & 
  \makecell{15.4$\pm$2.5 \\ (7.6)} & 
  \makecell{29.3$\pm$2.6 \\ (11.3)} & 
  \makecell{19.7$\pm$2.2 \\ (7.8)} \\ 
\addlinespace[2pt]
\textit{\textbf{full}} & 
  \makecell{27.3$\pm$2.6 \\ (16.6)} & 
  \makecell{20.2$\pm$2.5 \\ (12.4)} & 
  \makecell{28.9$\pm$2.6 \\ (10.2)} & 
  \makecell{19.5$\pm$2.2 \\ (7.6)} \\ 
\bottomrule
\end{tabular}
\end{table}

\begin{figure}[t!]
    \centering
    \includegraphics[width=\linewidth]{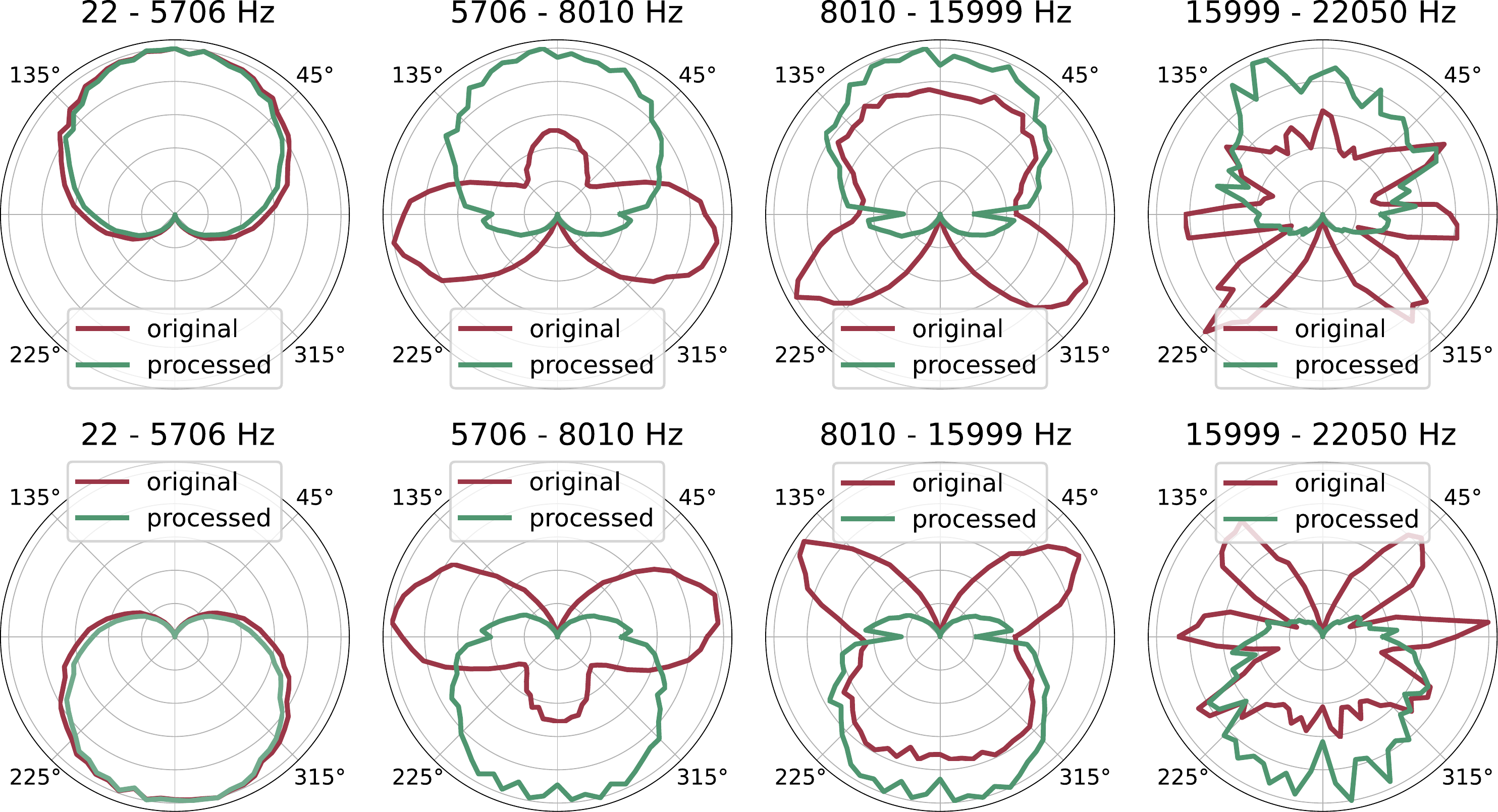}
    \caption{Spatial responses of the left- and right-facing cardioids of experiment \textit{\textbf{i) fix}}, respectively in the top and bottom row, in four frequency bands.}
    \label{fig:first_experiment}
\end{figure}

\begin{figure}[t!]
    \centering
    \includegraphics[width=\linewidth]{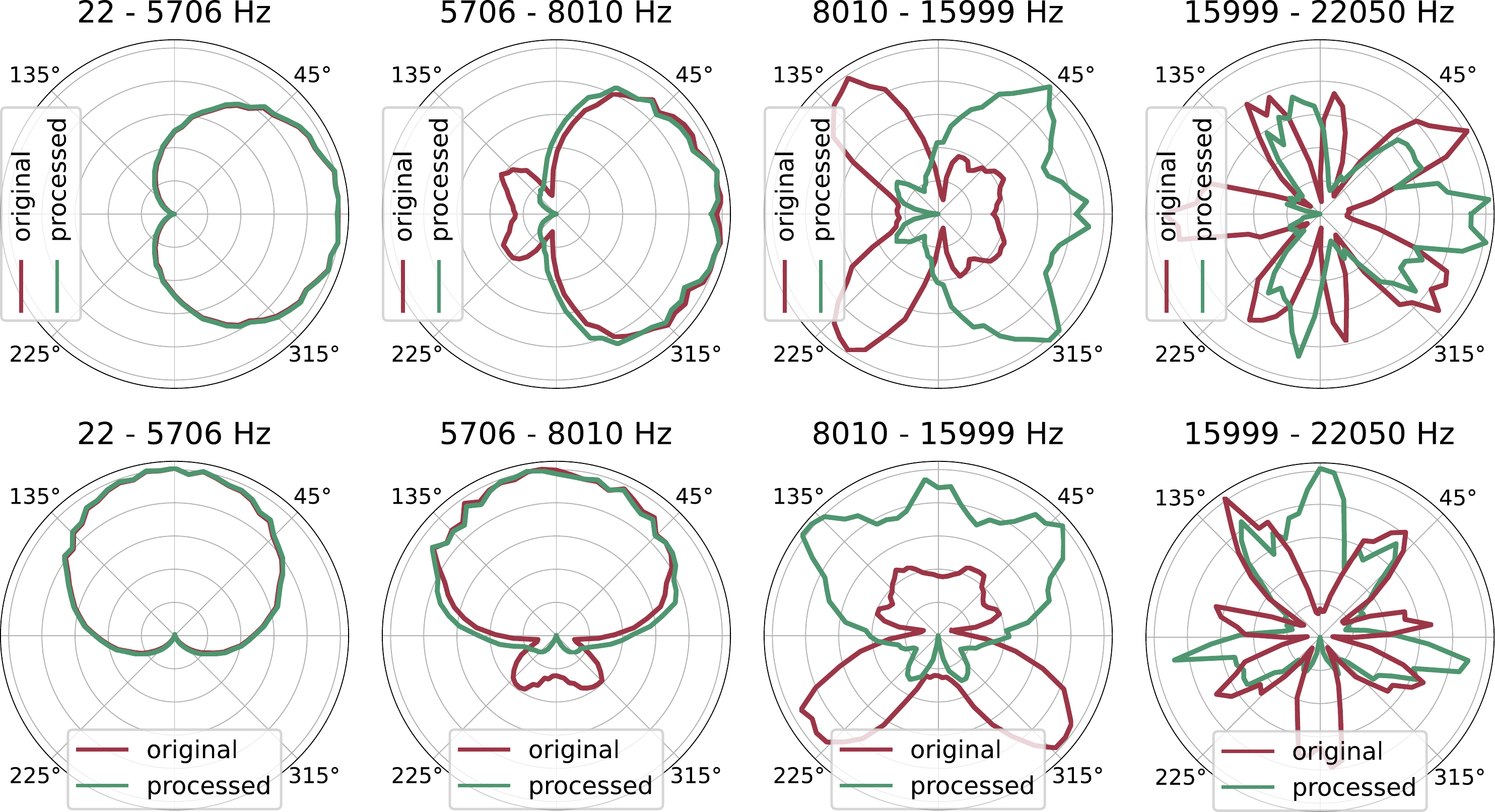}
    \caption{Spatial responses of the decoded front- and left-facing cardioids of experiment \textit{\textbf{ii) fix}}, respectively in the top and bottom row, in four frequency bands.}
    \label{fig:second_experiment}
\end{figure}

Figures \ref{fig:first_experiment} and \ref{fig:second_experiment} present polar plots of spatial responses, in case of the aliased and de-aliased virtual microphones, for experiments \textit{\textbf{i) fix}} and \textit{\textbf{ii) fix}}, respectively.
The patterns are shown in four frequency bands, the first of which is below \alsFreq{}.
Right- and rear-facing cardioids of experiment \textit{\textbf{ii)}} are symmetrical to their counterparts and were omitted for the sake of space.
In both experimental setups, the filters are effective in restoring the desirable spatial responses.
Even in the upper band, where the polar patterns are not accurately restored, the filters reduce the spurious components at the back and sides.

\subsection{Subjective evaluation}
\begin{figure}[t!]
    \centering
    \begin{subfigure}{0.23\textwidth}
        \includegraphics[width=\linewidth]{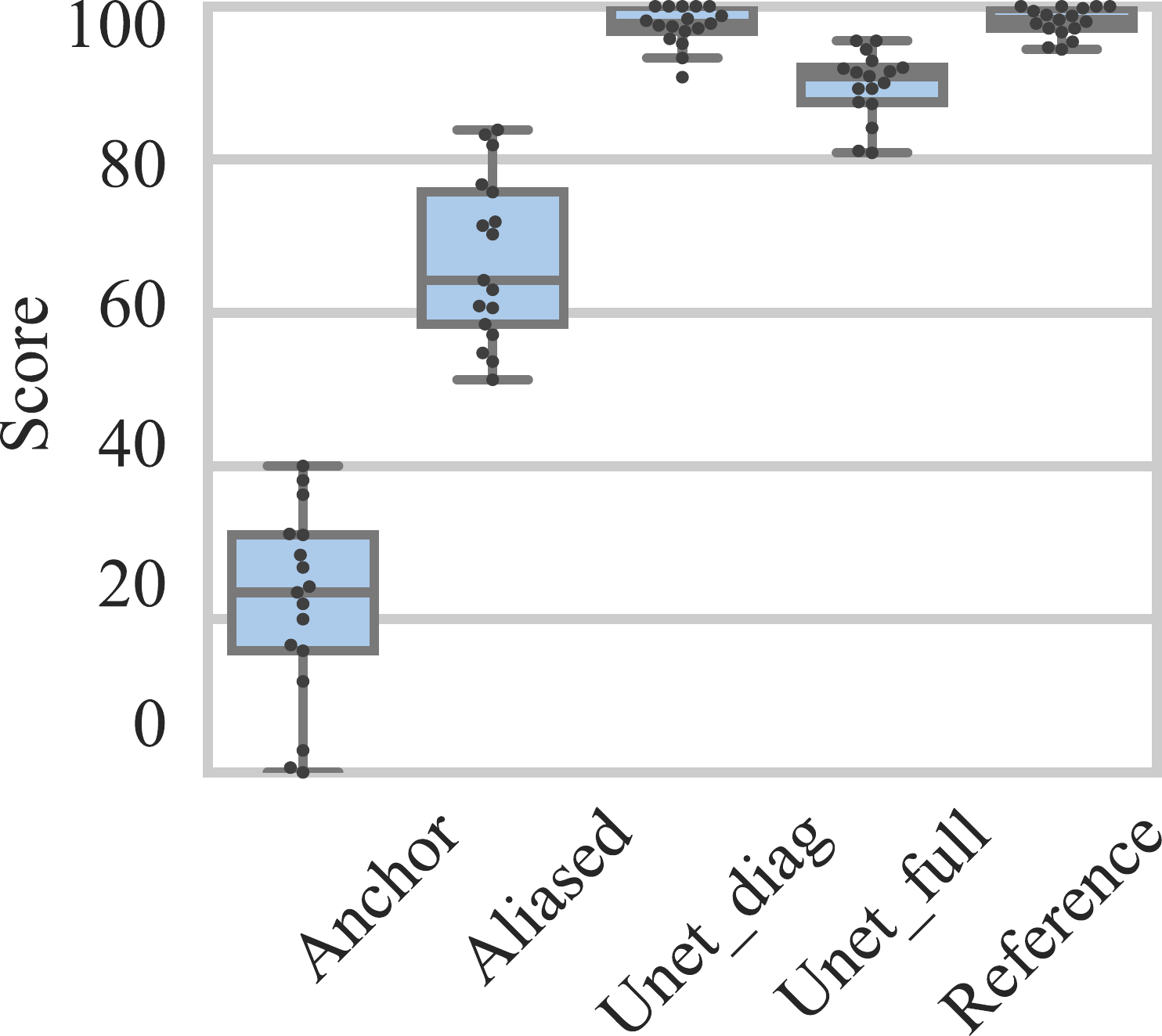}
        \caption{Timbre test}
    \end{subfigure}
    \hfill
    \begin{subfigure}{0.23\textwidth}
        \includegraphics[width=\linewidth]{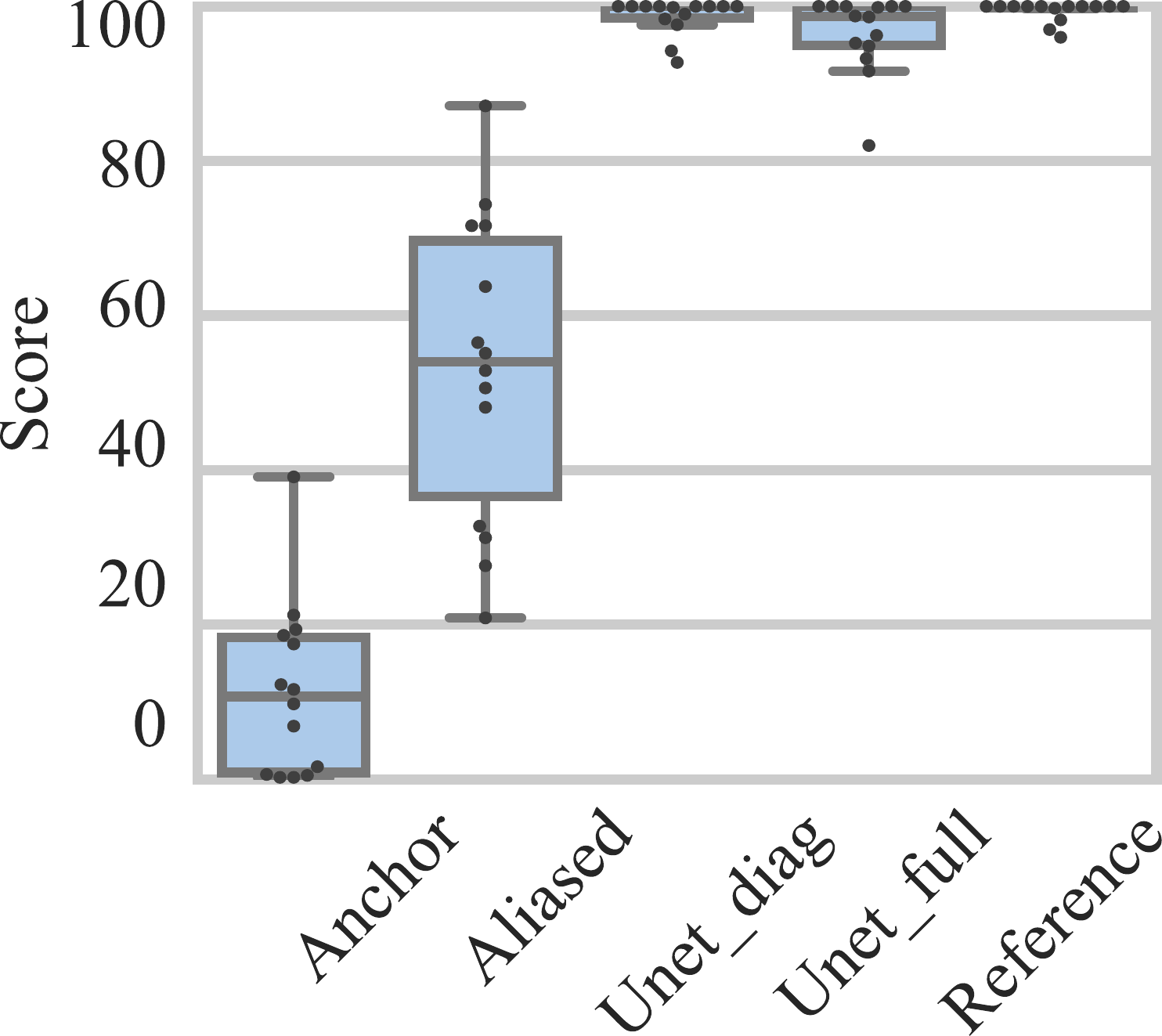}
        \caption{Localization test}
    \end{subfigure}
    \caption{Box plots for MUSHRA scores across different conditions and tests. Each point identifies the average rating for all test excerpts for one given listener.}
    \label{fig:mushra}
\end{figure}

To validate the perceptual impact of the proposed approach, we have designed two listening tests using the MUSHRA methodology \cite{mushra2014} and the webMUSHRA interface \cite{faucris.227085502}.
The first test studies the impact on the timbre of a  sound source with a random azimuth, decoded to mono using cardioid beamforming.
Subjects are presented with the reference alias-free signal, the aliased cardioid, an anchor low-passed \SI{3.5}{kHz}, and the de-aliased versions trained with the \textit{\textbf{diag}} or \textit{\textbf{full}} approaches, and are instructed to evaluate timbral similarity between each sample and the reference.
In the second test we study the ability to recover the  stereo localization by applying the networks to a pair of left- and right-facing cardioids. Here, a single sound source is positioned at a random azimuth between $\pm90^\circ$, encoded into the aliased and alias-free cardioids, and processed by the same networks, and subjects are instructed to evaluate the localization accuracy compared to the reference. The anchor is a version where the source is hard-panned to the left or right channel, whichever is more distant from the source position. Each test comprises nine excerpts of music, speech, or noise, evaluated by 17 expert listeners. The average ratings for all test excerpts were calculated per listener, with results presented in Figure \ref{fig:mushra}. 

A repeated measures ANOVA for the timbre test revealed a significant effect of rating stimuli ($p<0.001$).
Post-hoc tests\footnote{Multiple comparisons use the Benjamini–Hochberg procedure \cite{benjamini1995multiple}.} showed the aliased version scored 32 MUSHRA points below reference on average ($p<0.001$).
The \textit{\textbf{diag}} approach performed similarly to the reference (0.6 points below, $p=0.4$, not significant), while the \textit{\textbf{full}} approach scored significantly lower (9 points, $p<0.001$). Notably, a per-excerpt analysis revealed that with \textit{\textbf{full}} the difference was primarily due to a single broadband music item, scoring 53 MUSHRA points below reference ($p<0.001$), while other excerpts showed no significant differences to the reference (5  points at most,  $p>0.05$).

An analogous analysis for the localization test reveal a significant effect of rating stimuli ($p<0.001$). Post-hoc tests indicate the aliased version scored significantly lower than the reference (46 MUSHRA points below, $p<0.001$). Both the \textit{\textbf{diag}} and \textit{\textbf{full}} approaches perform similarly to the reference, with no significant differences (3 points at most, $p>0.05$).

These results demonstrate the effectiveness of the presented methods, in most cases rendering the timbre and localization indistinguishable from the reference.
The timbre test results align with the C-Si-SNR improvements, while the localization test confirms the successful restoration of polar patterns.
These patterns match the desired inter-channel level differences across most frequency bands, with only the highest bands showing deviations, which are perceptually less significant.
This comprehensive evaluation underscores the ability to effectively mitigate both the timbral and spatial spatial aliasing artifacts.

\section{Conclusions}
This article indicates the feasibility of restoring the timbre and localization degradations, caused by spatial aliasing, with a post-filtering neural network.
The framework has the flexibility to account for a decoder and a multi-channel setup, which lends itself to different applications.
The experiments indicate the ability to generalize across microphone arrays, while further research must include generalization across reverberation and diffuse noise.
Other interesting research possibilities include extension to frequency-dependent polar patterns and
feeding alias-free raw microphone signals to the network.


\clearpage

\section{Acknowledgments}
The authors would like to acknowledge Dr.~Xu Li, Dr.~Davide Scaini and Dr.~Chunghsin Yeh from Dolby Laboratories for fruitful discussions and the anonymous Interspeech reviewers for their valuable comments.

This work was supported by the National Science Centre, Poland, under grant number 2023/49/B/ST7/04100 and in part by the Excellence Initiative -- Research University Programme for the AGH University of Krakow. 
\bibliographystyle{IEEEtran}
\bibliography{references}

\end{document}